\documentclass{hep99}
\usepackage{epsf}

\newcommand{\dd}{{\rm d}}

\newcommand{\doverd}[1]{{\partial\over \partial #1}}
\newcommand{\oddo}[2]{#2{\dd #1\over \dd #2}}

\newcommand{\opo}{{\cal O}}
\newcommand{\msbar}{\overline{\rm MS}}
\newcommand{\nnb}{\nonumber}
\newcommand{\order}[1]{{\cal O}(#1)}
\newcommand{\bld}[1]{\boldmath{$#1$}}

\renewcommand{\Im}{{\rm Im}}
\newlength{\fsize}
\title{
  Quark mass effects to \bld{\sigma(e^+e^-\to \mbox{hadrons})}
  at \bld{\order{\alpha_s^3}}$^\dagger$
  \footnotetext[1]{Talk presented by R.H. at EPS HEP99, Tampere, Finland, July
  15-21, 1999.\\
  Work supported by DFG, Contract Ku 502/8-1.}
  \vspace{-6em}
  \begin{flushright}
    \normalsize
    \bf TTP99-42\\
    \bf BNL-HET-99/30\\
    \bf October 1999\\
    \bf hep-ph/9910345
  \end{flushright}
  \vspace{4em}
  }

\author{K.G. Chetyrkin$^1$, R. Harlander$^{1,2}$, and J.H. K\"uhn$^1$}
\address{$^1$University of Karlsruhe,
  D-76128 Karlsruhe, Germany\\ 
  $^2$Physics Department, Brookhaven National Laboratory, Upton, NY
  11973}

\abstract{ The calculation of the quartic mass terms at order
  $\alpha_s^3$ to the hadronic $R$ ratio is described. It is
  based on the operator product expansion for the quark current
  correlator, combined with an application of the renormalization group
  equation. }

\begin{document}
\maketitle

\mbox{}\thispagestyle{empty}\newpage
\mbox{}\thispagestyle{empty}\newpage
\mbox{}\thispagestyle{empty}\newpage

\title{Quark mass effects to \bld{\sigma(e^+e^-\to \mbox{hadrons})} 
  at \bld{\order{\alpha_s^3}}$^\dag$}
\maketitle
  \fntext{\dag}{
    Talk presented by R.H. at EPS HEP99.
    Work supported by DFG, Contract Ku 502/8-1.
    }
\setcounter{page}{1}
\thispagestyle{empty}

\section{Introduction}

The hadronic $R$ ratio is one of the most important quantities of QCD.
It is defined as the total cross section for hadronproduction
in $e^+ e^-$ annihilation, compared to the production rate for muon
pairs:
\begin{equation}
  R(s) = {\sigma(e^+e^-\to {\rm hadrons})\over \sigma(e^+e^-\to
    \mu^+\mu^-)}\,.
\end{equation}
Its inclusive character is a welcome feature both from the experimental
and the theoretical point of view. For example, the calculation of
$R(s)$ can be performed fully perturbatively, as long as bound state
effects which emerge close to threshold can be neglected.

\begin{figure}
\begin{center}
  \begin{tabular}{lr}
  \leavevmode \epsfxsize=\fsize \epsffile[160 265 420 470]{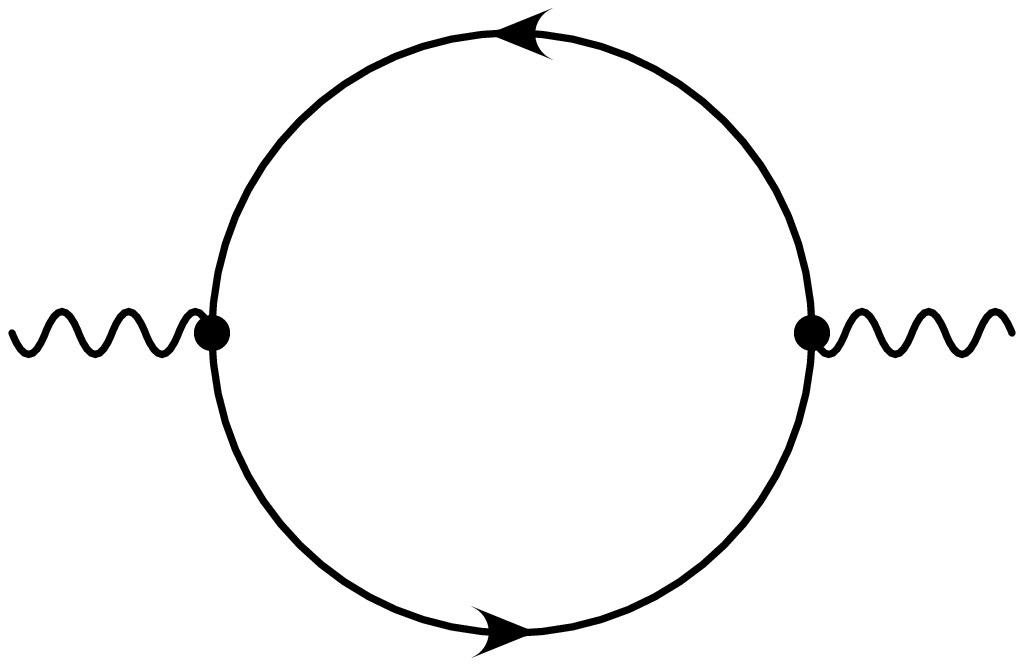} &
  \leavevmode \epsfxsize=\fsize \epsffile[160 265 420 470]{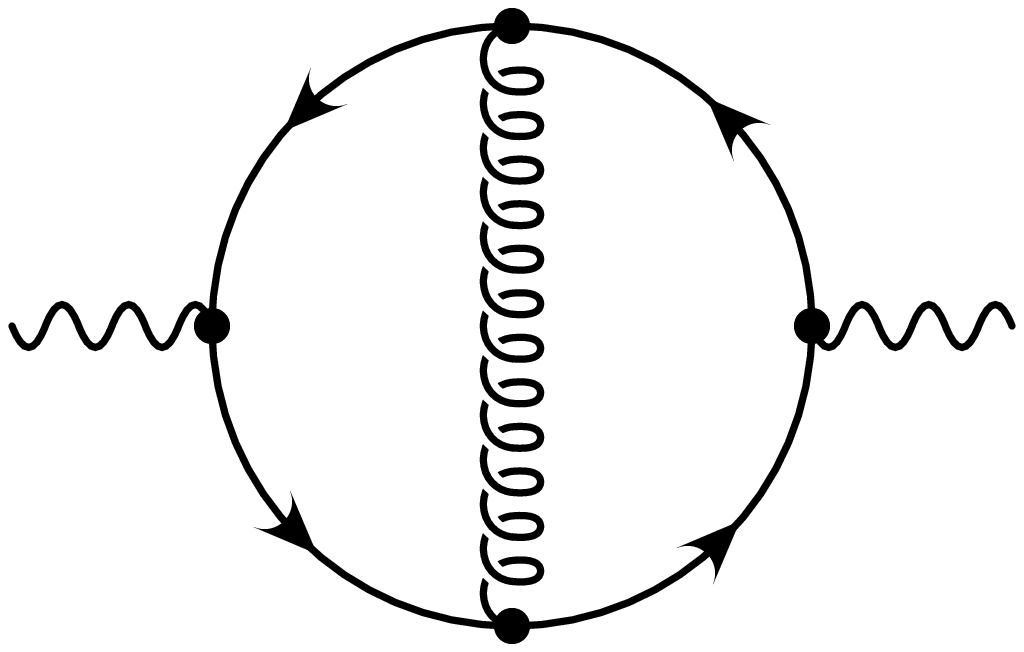}\\
  \leavevmode \epsfxsize=\fsize \epsffile[160 265 420 470]{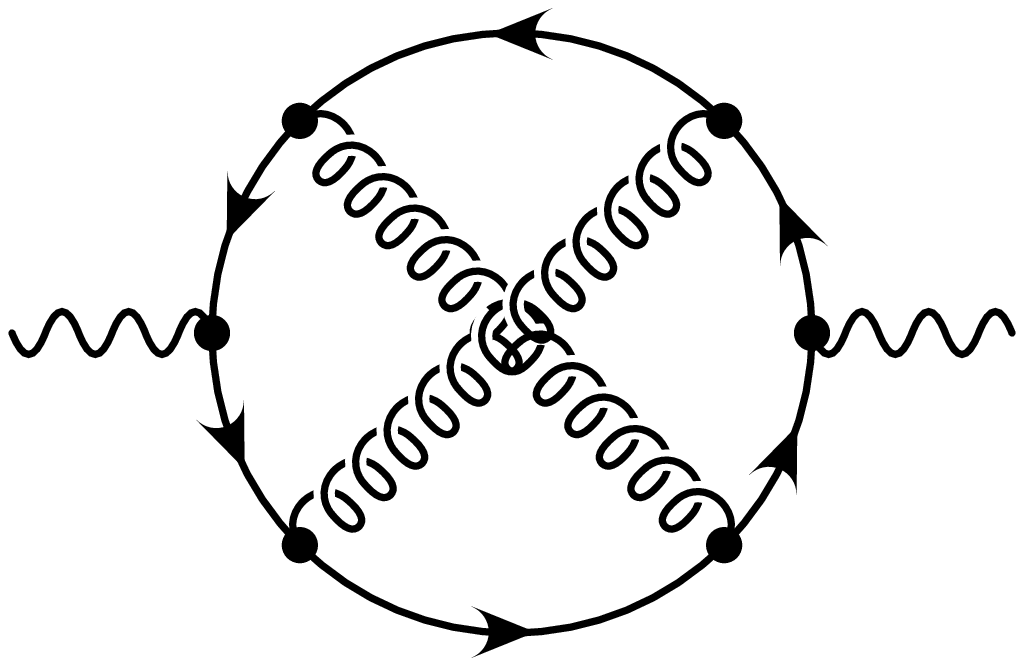} &
  \leavevmode \epsfxsize=\fsize \epsffile[160 265 420 470]{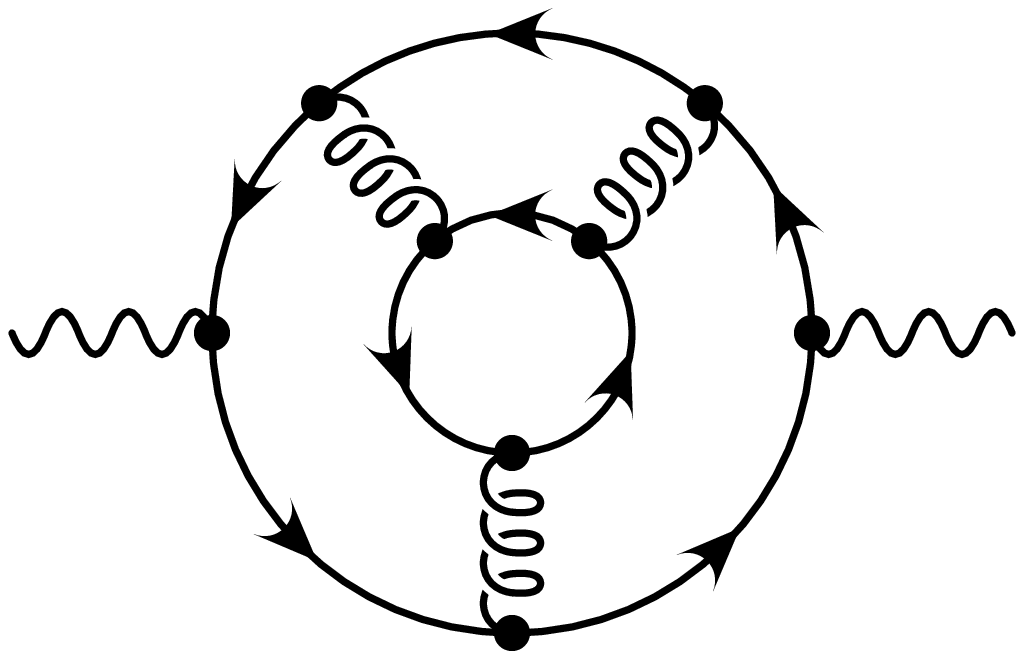}
\end{tabular}
\caption[]{\label{fig::pi} Sample diagrams contributing to $\Pi(q^2)$
  up to four-loop level.
  }
\end{center}
\end{figure}

The standard way to compute the hadronic $R$ ratio is to apply the
optical theorem which relates it to the imaginary part of the polarization
function:
\begin{equation}
R(s) = 12\pi\Im\Pi(s+i\epsilon)\,.
\label{eq::rimpi}
\end{equation}
Examples for diagrams contributing to $\Pi(q^2)$ up to order
$\alpha_s^3$ are shown in Fig.~\ref{fig::pi}. The analytic evaluation of
such diagrams is only possible up to two-loop order, if a non-vanishing
quark mass is assumed. At three-loop order, only the massless case can
be computed fully analytically. Nevertheless, mass effects at
$\order{\alpha_s^2}$ could be evaluated to very high accuracy, using two
different approximation methods. One is based on asymptotic expansions
\cite{smirnov}, yielding the result in terms of a series in $m^2/q^2$
times logarithmic terms of the form $\ln(-q^2/\mu^2)$ and
$\ln(-q^2/m^2)$, where $q^2$ is the renormalization scale
\cite{CHKS:m12}. The second method uses the moments of $\Pi(q^2)$,
defined as $C_n= (\partial^n/\partial (q^2)^n) \Pi(q^2)|_{q^2=0}$, to
reconstruct
$\Pi(q^2)$ in the full complex plane. The result for $R(s)$ is therefore
valid over the full energy range in this case \cite{pade}.

At $\order{\alpha_s^3}$, however, either of the two methods above
inevitably leads to four-loop diagrams whose evaluation is possible only
for a limited number of special cases at the moment. Nevertheless,
besides the massless approximation \cite{a3m0} which could be reduced to
a three-loop calculation using a method called infrared re-arrangement
\cite{IR},
also the quadratic mass corrections at $\order{\alpha_s^3}$ are
available \cite{a3m2}. They were obtained by a pure renormalization
group (RG) analysis using the terms at $\order{\alpha_s^2}$ as input
information.

At lower order in $\alpha_s$, the quartic ($m^4$) mass corrections turn
out to be quite sizable and are thus needed to provide a good
approximation to the full result over a reasonably large energy range.
This was the main motivation for us to evaluate these corrections at
$\order{\alpha_s^3}$. The method which was followed has been developed
in \cite{CK94}. At that time it was used to determine for the first time
the quartic corrections at $\order{\alpha_s^2}$. The strategy is to
evaluate the coefficient functions and the matrix elements of the
operator product expansion (OPE) for $\Pi(q^2)$ to (in our case) the
three-loop approximation and to use the RG equation in order to extend
the validity of the result to $\order{\alpha_s^3}$.

\section{Operator product expansion (OPE)}
\begin{figure}
\begin{center}
  \leavevmode \epsfxsize=\fsize \epsffile[180 265 400
  470]{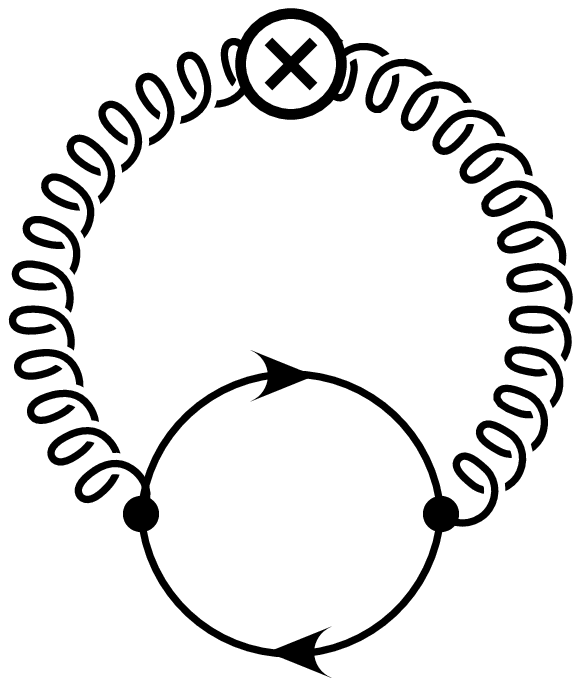}\quad 
  \epsfxsize=\fsize \epsffile[180 265 400
  470]{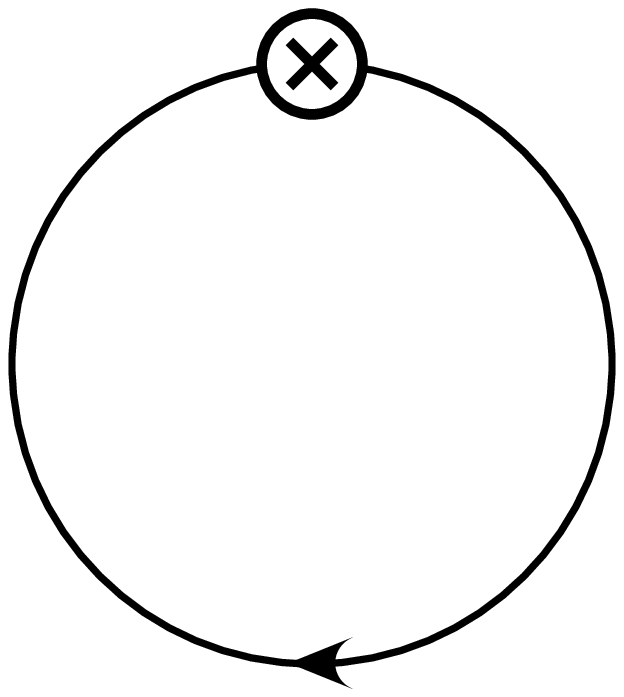}
\caption[]{\label{fig::vac} Sample diagrams contributing to $\langle
  \opo_1\rangle$ and $\langle\opo_2\rangle$.}
\end{center}
\end{figure}

The polarization function is defined as the vacuum expectation value of
the time ordered product of two currents, $j_\mu =
\bar\psi\gamma_\mu\psi$, where $\psi$ be a quark field with mass $m$.
In the limit of large $q^2$, $\Pi(q^2)$ can be written asymptotically as
an OPE:
\begin{eqnarray}
\Pi_{\mu\nu}(q) &=& i\int\dd^4 xe^{iq\cdot x}\langle 0|{\rm
  T}j_\mu(x)j_\nu(0)|0\rangle\nnb\\ 
&\stackrel{-q^2\to \infty}{\rightarrow}&
\sum_nC_{n,\mu\nu}\langle\opo_n\rangle\,,
\end{eqnarray}
where $\langle\opo_n\rangle\equiv \langle 0|\opo_n|0\rangle$.
Only Lorentz scalar operators contribute due to the external vacuum
states. The function $\Pi(q^2)$ in (\ref{eq::rimpi}) is defined as the
transversal part of $\Pi_{\mu\nu}(q)$,
\begin{equation}
\Pi(q^2) = {1\over q^2}{1\over D-1}\left(-g_{\mu\nu} + {q_\mu q_\nu\over
    q^2}\right)\Pi_{\mu\nu}(q)\,.
\end{equation}
An analogous definition is used to define transversal coefficient
functions $C_n$.  The operators up to dimension four are given by
\begin{eqnarray}
&&\opo^{(0)}= {\bf 1}\,,\qquad \opo^{(2)} = {\bf m^2}\,,
\label{eq::opo}
\\
&&\opo^{(4)}_1 = G_{\mu\nu}^2\,,\quad \opo^{(4)}_2 = m\bar\psi\psi\,,\quad
\opo^{(4)}_3 = {\bf m^4}\quad\nnb
\end{eqnarray}
(the superscript ``(4)'' will be dropped in what follows).
In addition to the renormalization of $m$ and
$\alpha_s$, one has to take into account operator mixing according to
\begin{equation}
\opo_n =\sum_m Z_{nm} \opo_m^{\rm B}\,.
\label{eq::ozob}
\end{equation}
In the $\msbar$ scheme which we will use throughout this paper,
operators of different mass dimension do not mix under renormalization.
For the dimension-4 operators the renormalization matrix $Z_{nm}$ has
been evaluated in \cite{CheSpi87}. It was expressed through the
renormalization constants for $m$ and $\alpha_s$, plus the one for the
QCD vacuum energy, $Z_0$. If no normal ordering is applied to the
operators, the perturbative expressions for their vacuum expectation
values are not necessarily zero. They lead to tadpole diagrams, examples
of which are shown in Fig.~\ref{fig::vac}. The matrix elements for the
``trivial'' operators are given by $\langle {\bf m}^{2n}\rangle =
m^{2n}$ to all orders of perturbation theory.  The ones for
$\opo_1$ and $\opo_2$ can be obtained up to
$\order{\alpha_s^2}$ by performing a three-loop calculation, as is
evident from a look at the diagrams in Fig.~\ref{fig::vac}.

\begin{figure}
  \begin{center}
  \begin{tabular}{cc}
    \epsfxsize=\fsize
    \epsffile[130 250 450 460]{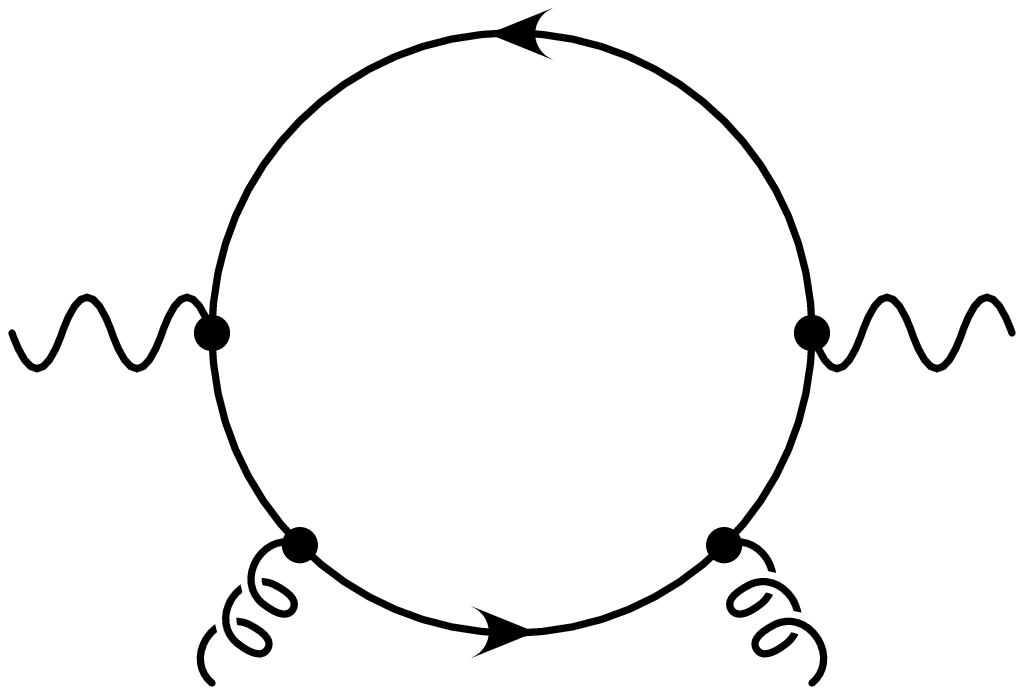} &
    \epsfxsize=\fsize
    \epsffile[130 250 450 460]{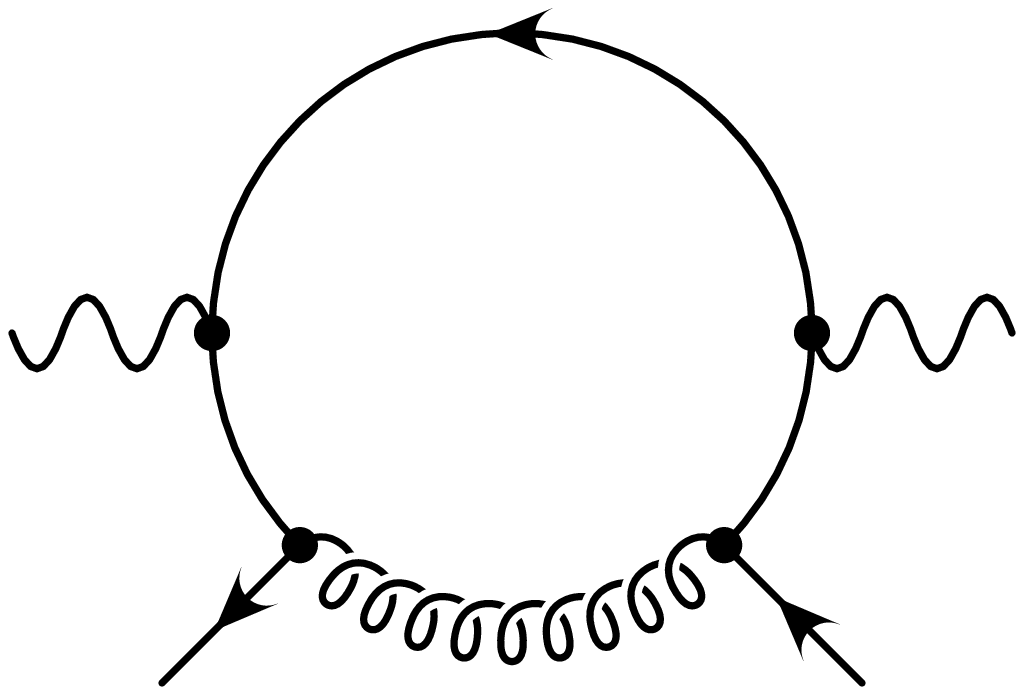}\\
    \multicolumn{2}{c}{
    \epsfxsize=\fsize
    \epsffile[130 250 450 460]{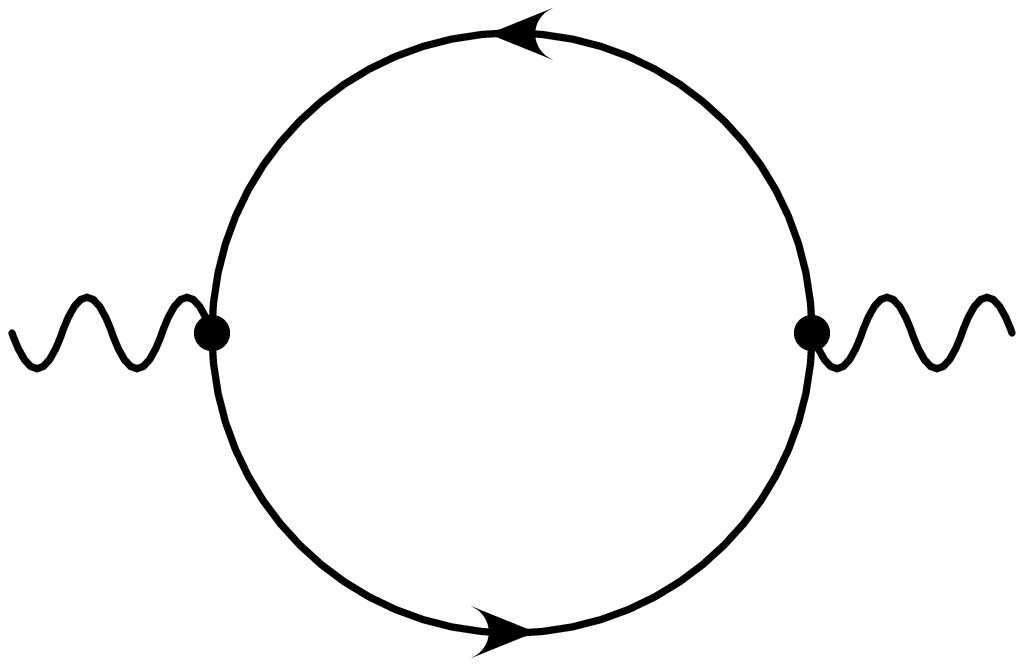}}
  \end{tabular}
\caption[]{\label{fig::cs} Sample diagrams contributing to $C_1$,
  $C_2$, and $C_3$.}
\end{center}
\end{figure}

Concerning the coefficient functions it has been shown in \cite{CheSpiGor85}
that they depend only on the external momentum if minimal subtraction is
used as the renormalization scheme. The most elaborate method to reduce
their evaluation to massless propagator diagrams is called the ``method
of projectors'' \cite{refproj}. It leads to diagrams like the ones
shown in Fig.~\ref{fig::cs}, where it is understood that certain
derivatives w.r.t.\ the quark mass and the momentum carried by the
external quark and gluon lines are to be applied to the integrand.
Before integration, however, all masses and external momenta (except
$q$) can be set to zero. A glance at these diagrams shows that a
three-loop calculation yields $C_1$ and $C_2$ with $\order{\alpha_s^3}$,
but $C_3$ only with $\order{\alpha_s^2}$ accuracy. Furthermore, one
realizes that the imaginary parts arising from $C_1$ and $C_2$ start to
be non-zero only at $\order{\alpha_s}$. Thus the $\alpha_s^2$
calculation for the matrix elements mentioned before is sufficient to
determine the imaginary part of the products $C_1\langle\opo_1\rangle$
and $C_2\langle\opo_2\rangle$ to $\order{\alpha_s^3}$.

The crucial step to reconstruct the $\alpha_s^3$ terms of $C_3$ is now
to employ RG invariance of the dimension-4 piece of the OPE:
\begin{equation}
\oddo{}{\mu^2}\sum_{n=1}^3 C_n\opo_n = 0\,.
\label{eq::ope}
\end{equation}
The effect of $\oddo{}{\mu^2}$ on the $\opo_n$ is determined by
(\ref{eq::ozob}):
\begin{equation}
\oddo{}{\mu^2}\opo_n = \sum_m\gamma_{nm}\opo_n\,,
\end{equation}
where $\gamma_{mn}$ is the anomalous dimension of the operators which is
related to the renormalization matrix $Z_{mn}$ in the standard way.
For the coefficient functions, the derivative w.r.t.\ $\mu$ will be
rewritten as
\begin{equation}
\oddo{}{\mu^2} C_n = \left(\doverd{L} +
  \alpha_s\beta\doverd{\alpha_s}\right)C_n\,,
\end{equation}
with the QCD $\beta$ function and $L=\ln(-\mu^2/q^2)$.
Solving (\ref{eq::ope}) for $\doverd{L}C_3$ and keeping only terms
proportional to $\opo_3$, one obtains
\begin{equation}
\doverd{L}C_3 = - \alpha_s\beta\doverd{\alpha_s}C_3 - 4\gamma^m C_3 -
\sum_{n=1}^2 \gamma_{n3} C_n\,,
\label{eq::CK}
\end{equation}
where $\gamma_{33} = 4\gamma^m$ has been used ($\gamma^m$ is the quark
anomalous dimension).  Since $\beta$ and $\gamma^m$ do not carry terms
of $\order{\alpha_s^0}$, the r.h.s.\ of (\ref{eq::CK}) is known to
$\order{\alpha_s^3}$, and so is the l.h.s. The logarithmic terms (and
thus the imaginary part) of $C_3$ may therefore be obtained by trivial
integration.

\section{Results}
Let us consider the case of charm quark pair production.  The general
result for the terms of order $\alpha_s^3 m^4/s^2$ to $R(s)$ will be
given elsewhere \cite{CHK:prep}. The upper limit of the energy region we
are interested in is dictated by the $b\bar b$ threshold.  The lower one
is such that, on the one hand, the massless charm approximation fails,
and on the other hand, one is sufficiently far above the $c\bar c$
threshold in order for the small mass expansion to be valid (see
\cite{CHKS:m12} for a detailed analysis on the convergence properties at
$\order{\alpha_s^2}$).
\begin{figure}
  \begin{center}
    \leavevmode
    \epsfxsize=6.cm
    \epsffile[110 265 465 560]{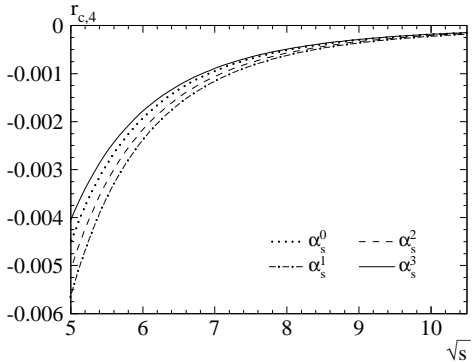}
    \hfill
    \caption[]{\label{fig::rvc4}\sloppy
      Quartic mass corrections to charm pair production.
      }
  \end{center}
\end{figure}
Fig.~\ref{fig::rvc4} shows the
result for the contribution where the external photon couples directly
to the charm quark. There is also a contribution arising from diagrams
with only massless quarks coupling to the photon and the charm quarks
emerging only through their coupling to gluons. In turns out, however,
that the latter piece is tiny, so we will neglect it here.
The notation is given by the following decomposition:
\begin{equation}
R_c(s) = 3\,\Big[r_{c,0}(s) + r_{c,2}(s) + r_{c,4}(s)\Big] + \ldots\,,
\end{equation}
which means that $r_{c,0}(s)$ corresponds to the massless
approximation and $r_{c,2}(s)$ and $r_{c,4}(s)$ contains the
mass corrections of order $m^2$ and $m^4$, respectively.

The charm quark mass is taken in the $\msbar$ scheme and the
renormalization scale is set to $\mu^2=s$. One observes that the higher
order terms in $\alpha_s$ are all of comparable magnitude, and the
convergence of the perturbative series is not manifest. Compared to the
Born result, however, the corrections in $\alpha_s$ are small such that
the stability of the full prediction is almost unaffected by the higher
order terms.

\def\app#1#2#3{{\it Act.~Phys.~Pol.~}{\bf B #1} (#2) #3}
\def\apa#1#2#3{{\it Act.~Phys.~Austr.~}{\bf#1} (#2) #3}
\def\cmp#1#2#3{{\it Comm.~Math.~Phys.~}{\bf #1} (#2) #3}
\def\cpc#1#2#3{{\it Comp.~Phys.~Commun.~}{\bf #1} (#2) #3}
\def\epjc#1#2#3{{\it Eur.\ Phys.\ J.\ }{\bf C #1} (#2) #3}
\def\fortp#1#2#3{{\it Fortschr.~Phys.~}{\bf#1} (#2) #3}
\def\ijmpc#1#2#3{{\it Int.~J.~Mod.~Phys.~}{\bf C #1} (#2) #3}
\def\ijmpa#1#2#3{{\it Int.~J.~Mod.~Phys.~}{\bf A #1} (#2) #3}
\def\jcp#1#2#3{{\it J.~Comp.~Phys.~}{\bf #1} (#2) #3}
\def\jetp#1#2#3{{\it JETP~Lett.~}{\bf #1} (#2) #3}
\def\mpl#1#2#3{{\it Mod.~Phys.~Lett.~}{\bf A #1} (#2) #3}
\def\nima#1#2#3{{\it Nucl.~Inst.~Meth.~}{\bf A #1} (#2) #3}
\def\npb#1#2#3{{\it Nucl.~Phys.~}{\bf B #1} (#2) #3}
\def\nca#1#2#3{{\it Nuovo~Cim.~}{\bf #1A} (#2) #3}
\def\plb#1#2#3{{\it Phys.~Lett.~}{\bf B #1} (#2) #3}
\def\prc#1#2#3{{\it Phys.~Reports }{\bf #1} (#2) #3}
\def\prd#1#2#3{{\it Phys.~Rev.~}{\bf D #1} (#2) #3}
\def\pR#1#2#3{{\it Phys.~Rev.~}{\bf #1} (#2) #3}
\def\prl#1#2#3{{\it Phys.~Rev.~Lett.~}{\bf #1} (#2) #3}
\def\pr#1#2#3{{\it Phys.~Reports }{\bf #1} (#2) #3}
\def\ptp#1#2#3{{\it Prog.~Theor.~Phys.~}{\bf #1} (#2) #3}
\def\ppnp#1#2#3{{\it Prog.~Part.~Nucl.~Phys.~}{\bf #1} (#2) #3}
\def\sovnp#1#2#3{{\it Sov.~J.~Nucl.~Phys.~}{\bf #1} (#2) #3}
\def\tmf#1#2#3{{\it Teor.~Mat.~Fiz.~}{\bf #1} (#2) #3}
\def\yadfiz#1#2#3{{\it Yad.~Fiz.~}{\bf #1} (#2) #3}
\def\zpc#1#2#3{{\it Z.~Phys.~}{\bf C #1} (#2) #3}
\def\ibid#1#2#3{{ibid.~}{\bf #1} (#2) #3}

\end{document}